\documentstyle[aps]{revtex}
\begin{document}
\draft
\title{Effect  of  viscosity on one dimensional hydrodynamic flow
and direct photons from 200 AGeV S+Au collisions at CERN SPS.}

\author{\bf A. K. Chaudhuri\cite{byline}}

\address{ Variable Energy Cyclotron Centre\\
1/AF,Bidhan Nagar, Calcutta - 700 064\\}

\maketitle

\begin{abstract}
We  have  analysed  the  direct  photon data obtained by the WA80
collaboration in 200 A GeV S+Au collision at CERN SPS, in  a  one
dimensional  hydrodynamical  model.  Two scenario was considered:
(i) formation of quark-gluon plasma and  (ii)  formation  of  hot
hadronic  gas.  For both the scenario, ideal as well as extremely
viscous fluid was considered. It was  found  that  direct  photon
yield  from QGP is not affected much whether the fluid is treated
as  ideal  or  extremely  viscous.  The  yield   however   differ
substantially if hadron gas is produced. Both the scenario do not
give satisfactory description of the data. \end{abstract}

\pacs{ PACS numbers(s):12.38.Mh,24.85.+p,13.85.Qk,25.75.+r}

Recently  WA80  collaboration  has  published their direct photon
emission  data  for  200  A  GeV  S+Au  collisions  at  CERN  SPS
\cite{al96}.  The  final  version  differ  from their preliminary
analysis \cite{sa93} in several ways. The data  was  improved  by
increasing  the  detector  coverage  and  also enlarging the data
sample. The analysis  technique  was  also  improved  upon.  Most
importantly,  the data now gives the upper limit of the invariant
yield for direct photon production at the 90\% confidence  limit,
rather  than  the absolute yield. Much interest was aroused after
the publication of the WA80  preliminary  result  of  the  direct
photon  data \cite{sa93}, as it was hoped that the direct photons
can be a conclusive probe of the much debated quark-gluon  plasma
(QGP)   expected   to  be  produced  in  relativistic  heavy  ion
collisions.  The  preliminary  data  were  analysed  by   several
authors. Xiong and Shuryak \cite{sh94} analysed the data assuming
a  mixed phase formation and found excess photons. Srivastava and
Sinha \cite{sr94} analysed  the  data  considering  two  possible
scenarios  after  the collision, one with the phase transition to
QGP, the other without it. It was  claimed  that  the  data  were
explained  only  in  the  phase  transition scenario. We had also
analysed the preliminary version of the WA80 direct  photon  data
\cite{ch95}.  It was shown that formation of {\em viscous} hadron
gas in the initial state, can explain the data.

In  the  present  paper,  we analyse the WA80 direct photon data,
assuming formation of QGP or hot  hadronic  gas  in  the  initial
state.  In cognizance of the fact that QGP or hot hadronic matter
are not ideal, we consider two extreme cases of fluid  flow,  (i)
ideal  fluid flow and (ii) maximal viscous flow. Reaslistic fluid
flow must be within  this  two  extreme  limits.  Though,  photon
spectra  thus  calculated,  will  have  uncertainties, as will be
shown below,  important  conclusions  can  still  be  reached  by
comparing with experimental data.

Here  we consider two possible scenarios that can arise after the
collision: (i) the  phase  transition  (PT)  scenario,  when  the
collision  leads  to  formation of baryon free quark-gluon plasma
(QGP) and (ii) the no phase transition scenario (NPT),  when  hot
hadronic  gas,  again baryon free, is formed after the collision.
In the PT scenario, it is assumed that  after  the  collision,  a
baryon  free  QGP  is  formed  at  initial  time  $\tau_i$ and at
temperature  $T_i$.  It  expands  and  cools  till  the  critical
temperature  $T_c$ is reached and the fluid enters into the mixed
phase (we are assuming a 1st order phase transition). The  matter
remains  in the mixed phase until all the QGP matter is converted
adiabatically into the hadronic matter. The hadronic matter  then
cools further till the freeze-out temperature $T_F$ is reached at
time  $\tau_F$.  In  the NPT scenario it will be assumed that the
matter is formed as (baryon free) hot  hadronic  gas  at  initial
time $\tau_i$ at temperature $T_i$. It expands and cools till the
freeze-out  temperature $T_F$ is reached at $\tau_F$. In both the
scenario, as mentioned earlier, we will consider  the  ideal  and
the   maximal   viscous   fluid   flow.  We  assume  longitudinal
boost-invariant hydrodynamic expansion  of  the  system,  without
transverse  expansion. We feel this to be justified at SPS energy
as transverse expansion is small \cite{ja93} and  its  effect  on
photon production is only marginal \cite{sr94}.

The  space-time  evolution  of the fluid (QGP or hadronic gas) is
governed by the energy-momentum conservation equation, which  for
a    longitudinal    similarity    flow   can   be   written   as
\cite{da85,ho85,bj82},

\begin{equation}
\frac{d\varepsilon}{d\tau} =
-(\varepsilon +p-\frac{4\eta}{3\tau}-\frac{\zeta}{\tau})/\tau
\label{1}
\end{equation}

\noindent  where,  $\eta$ and $\zeta$ are the shear viscosity and
the bulk viscosity coefficients. The other variables  are  having
their  usual  meaning.  In the ideal case, $\eta=\zeta=0$. In the
other    extreme    case    with    maximal    viscous     force,
$(4\eta/3+\zeta)/\tau=p$,  as  the total viscous force can not be
larger than the pressure. For both the cases eqn.\ref{1}  can  be
solved  analytically for a $p=1/3\varepsilon$ type of equation of
state. In the present work, for the QGP phase the bag equation of
state $p_q=a_qT^4-B$, with $a_q=37\pi^2/90$ and for the  hadronic
phase  the  pion gas equation $p_h=a_hT^4$ with $a_h=4.6\pi^2/90$
were used\cite{sr94,ch95}. The mixed phase was described  by  the
Maxwell  construct  $p_q(T_c)=p_h(T_c)$, which also gives the bag
constant B.

For  the ideal fluid, as the expansion is isentropic, the initial
temperature $T_i$ of the fluid at an initial time $\tau_i$ can be
obtained by relating the entropy density with the  observed  pion
multiplicity   (assuming   pion   decoupling   to  be  adiabatic)
\cite{hw85},

\begin{equation}
T^3_i\tau_i=\frac{1}{\pi R^2_A} \frac{c}{4a_i}\frac{dN}{dy}
(b=0) \label{2a}
\end{equation}

\noindent  where  $c=2\pi^4/45\zeta(3)$,$a_i=g_{q,h}\pi^2/90$ and
$R_A$ is the transverse radius of the system.  $b=0$  corresponds
to  central  collisions.  However,  for  viscous flow, entropy is
generated, the flow is no longer isentropic, the  above  equation
can  not  be  used.  However,  we  can  still  assume  that  pion
decoupling is adiabatic. Since bulk of the pions are produced  at
freeze-out,  as argued in \cite{ch95} we equate the final entropy
density with the pion multiplicity.

\begin{equation}
T^3_f\tau_f=\frac{1}{\pi R^2_A} \frac{c}{4a_h}\frac{dN}{dy}
(b=0) \label{2b}
\end{equation}

\noindent  to  obtain  the  freeze-out  time $\tau_F$ for a given
freeze-out temperature $T_F$.  The  initial  temperature  of  the
fluid  is then obtained from eq.\ref{1}, for a given $\tau_i$. In
both the PT and NPT scenarios, the initial time i.e.  the  proper
time   from  which  onward  the  hydrodynamical  approach  become
applicable, is not known. It is presumed  to  be  small  and  one
generally  uses the canonical value 1 fm. However, $\tau_i$'s for
the  QGP  and  the  hadron  gas  need  not  be  same.  Elementary
considerations suggests that $\tau_i$ for the hadronic gases will
be  three  times  larger  than  that  for the QGP \cite{ch95}. In
cognizance of the uncertainty in $\tau_i$,  we  have  used  three
different  values:  $\tau_i$=0.3,0.6  and 1 fm in the PT scenario
and $\tau_i$=1,2 and 3 fm in the NPT scenario.  In  table  1,  we
have shown the initial temperature of the fluid in the PT and NPT
scenario  for a pion multiplicity dN/dy=225, which is appropriate
for the  central  S+Au  collisions  \cite{al92}.  The  transition
temperature  for  the  QGP  fluid was fixed at $T_c$=160 MeV. The
freeze-out temperature was taken as $T_F$=100 MeV.  In  both  the
scenario,  the  initial temperature differ quite substantially in
case of ideal and maximal viscous flow.  Ideal  fluid  is  always
formed  at  (20-30 \%) higher temperature than the viscous fluid.
This is because entropy is generated during the viscous flow.

Direct  photon  yield  was  obtained  by  convoluting the rate of
emission of photons with the  space-time  evolution,  using  well
established methods \cite{sr94,ch95,ja93}. For the thermal photon
production  rate  from  an equilibrated hadron gas, consisting of
$\pi$,$\rho$, $\eta$ and $\omega$, we  use  the  parametric  form
given by Kapusta et al \cite{ka91}.

\begin{equation}
E\frac{dR^\gamma}{d^3p}=\frac{5\alpha\alpha_s}{18\pi^2} T^2 e^{E/T}
\ln[1+\frac{2.912E}{g^2T}] \label{3}
\end{equation}

\noindent  where E is the photon energy in the local rest frame .
In the following we fix $\alpha_s  =g^2/4\pi=0.4$.  Incidentally,
eqn.\ref{3}  also  describes the direct photon emission rate from
QGP. In a hadronic gas at temperature T>100 MeV, a $\pi\rho$ pair
can easily form an $A_1(1260)$  resonance,  which  can  decay  to
photon  \cite{xi92}. This process is not included in eqn.\ref{3}.
We have included  the  direct  photon  emission  from  the  $A_1$
resonance  using  the  parametric  form  given  by  Xiong  et  al
\cite{xi92},

\begin{equation}
E \frac{dR}{d^3p}= 2.4 \times T^{2.15} exp[-1/(1.35E)^{0.77}-E/T]
(fm^{-4}GeV^{-2}) \label{4}
\end{equation}

In  fig.1,  we  have shown the WA80 revised data (the solid bars)
and direct photon yield in the PT scenario (the  shaded  region).
The  shaded  region indicate the uncertainty in the direct photon
production due to viscosity, the upper and lower limits of  which
were   obtained  by  assuming  ideal  and  maximal  viscous  flow
respectively. The uncertainty in direct photon yield is not much,
though the initial temperature  of  the  ideal  and  the  maximal
viscous flow differ substantially. This can be understood. In the
QGP  scenario,  both for the ideal and the viscous flow, the bulk
of the photons are from the mixed and the hadron phase where  the
temperature  variation  is  limited  from  160-100  MeV.  Then as
viscosity is proportional to temperature, effect of viscosity  is
less  in  QGP.  {\em  For  all  practical  purposes, fluid can be
considered ideal in the PT scenario}. It is interesting  to  note
that  photon spectra are nearly identical whether $\tau_i$=0.3 fm
or 1.0 fm, though  the  initial  temperatures  are  substantially
different.  This  indicates  that  contribution to direct photons
from pure  QGP  sector  is  small  compared  to  hadronic  scetor
\cite{sr94}.  It  is  also  evident that the PT scenario does not
describe the WA80 data well. Insufficient number of  large  $p_T$
photons  are produced and the data are underpredicted by a factor
of 10 or more at large $p_T$.

In  fig.2,  we  have  shown the same results in the NPT scenario.
Here again, the shaded region indicate  the  uncertainty  due  to
viscosity.  In contrast to the PT scenario, we find a substantial
difference in  the  photon  yield  with  $\tau_i$,  As  mentioned
earlier,  in the PT scenario, initial QGP phase contribute little
to the photon yield. On  the  otherhand,  in  the  NPT  scenario,
contributions  from  the  initial  state are significant. Then as
$\tau_i$ is increased, photon yield at high $p_T$ is reduced,  as
a  result  of  reduced  initial temperature. For the same reason,
uncertainty  in  direct  photon  yield  due   to   viscosity   is
substantial  in  this  scenario.  The uncertainty also grows with
$p_T$. This is also understood easily. Initially temperatures  of
the  ideal  and extreme viscous fluid differ quite substantially,
however, at later time this difference  is  reduced,  ultimately,
both  the  fluid  freeze-out  at the same temparature at the same
time. Accordingly, uncertainty is  large  at  high  $p_T$  (early
times)  and  lessened  at  low $p_T$ (later times). It is evident
that if the collision leads to formation of ideal hadron gas, the
data  is  not  explained.  More  photons  are  produced  than  in
experiment.  For  $\tau_i$=1  fm,  even extreme viscosity can not
lower the yield to agree with experiment.  For  $\tau_i$=2-3  fm,
the  large  $p_T$  photons  are within the shaded region, but the
data are overpredicted in the intermediate (1-2 GeV) $P_T$ range.
The NPT scenario also do not give satisfactory description of the
WA80 direct photon data.

One  of  the  deficiency  of  the present model is the neglect of
pre-equilibrium photons  emitted  before  $\tau_i$.  Traxler  and
Thoma  \cite{tr96}  calculated  pre-equilibrium photons and found
them to be order of magnitude less than the equilibrium  photons.
Recently  Roy  et  al \cite{pr97} also calculated pre-equilibrium
photons.  They  used  Fokker-Plank  equations  and   found   that
pre-equilibrium  photons  are  less  or  at  best  equal  to  the
equilibrium  photons.  Even  if   the   pre-equilibrium   photons
contribute  as  much  as  equilibrium photon, in the PT scenario,
their inclusion will not raise the calculated yield  sufficiently
high  to agree with experiment. In the NPT scenario, inclusion of
pre-equilibrium photons can only worsen the descrepency  further.
The  present result that both the PT and NPT scenario do not give
satisfactory description of the  WA80  direct  photon  data  will
remain  valid  even if pre-equilibrium photons are included. Some
unconventional scenario is then needed to explain the WA80 direct
photon data. For example, in the PT scenario, as argued by  Xiong
and Shuryak \cite{sh94}, if the fluid expands slowly in the mixed
phase,  (very  soft  equation  of state), the photon yield can be
increased.

In  conclusion, we have studied the effect of viscosity on direct
photon production in PT and NPT scenario, currently in vouge.  In
the  PT  scenario,  when  QGP  is  assumed  to be produced in the
initial state, it was shown  that  the  effect  of  viscosity  on
direct  photon  yield is minimum. Fluid in the PT scenario can be
well approximated by a ideal fluid. This is not  so  in  the  NPT
scenario.  Viscosity  can  have  a  large effect there, and it is
essential that one take into the account of viscosity in the  NPT
scenario.  Though,  the present calculations were done for direct
photons, these conclusions  are  valid  for  other  probes  also.
Viscosity coefficients of hot hadronic matter are then of primary
importance,   and   efforts   must  be  made  to  calculate  them
accurately. The present study also shows that the revised version
of WA80 direct photon data could not be explained in  the  PT  or
the NPT scenario, irrespective of the viscosity.

\begin{figure}
\caption{ The direct photon yield in the phase transition scenario
for  three  different  $\tau_i$'s.  The  solid  bars are the WA80
direct photon data.}
\end{figure}
\begin{figure}
\caption{The  direct  photon  yield  in  the  no  phase transition
scenario for three different $\tau_i$'s. The solid bars  are  the
WA80 direct photon data.}
\end{figure}

\begin{table}
\caption{The initial  temperature of the ideal fluid $T^{ideal}_i$
and the extremely viscous fluid $T^{viscous}_i$   in  the  phase
transition  (QGP)  and  the  no  phase  transition  (Hadronic  gas)
scenario for different initial time $\tau_i$'s.
\label{table1}}
\begin{tabular}{ccccccc}
\multicolumn{3}{c}{QGP} &&\multicolumn{3}{c}{Hadronic Gas}\\
$\tau_i$ & $T^{ideal}_i$ & $T^{viscous}_i$ &&$\tau_i$& $T^{ideal}_i$ &
$T^{viscous}_i$\\
(fm)&(MeV)&(MeV)&&(fm)&(MeV)&(MeV)\\
\tableline
0.3 & 304 & 226 && 1.0 & 407 & 287\\
0.6 & 241 & 187 && 2.0 & 323 & 241\\
1.0 & 203 & 160\tablenote{For $\tau_i$=1fm, the multpilicity is not enough
to  produce the fluid in pure QGP state. We then assumed it to be
formed in the  mixed  phase  with  proper  fraction  of  QGP  and
hadronic fluids.}
&& 3.0 & 283 & 218\\
\end{tabular}
\end{table}
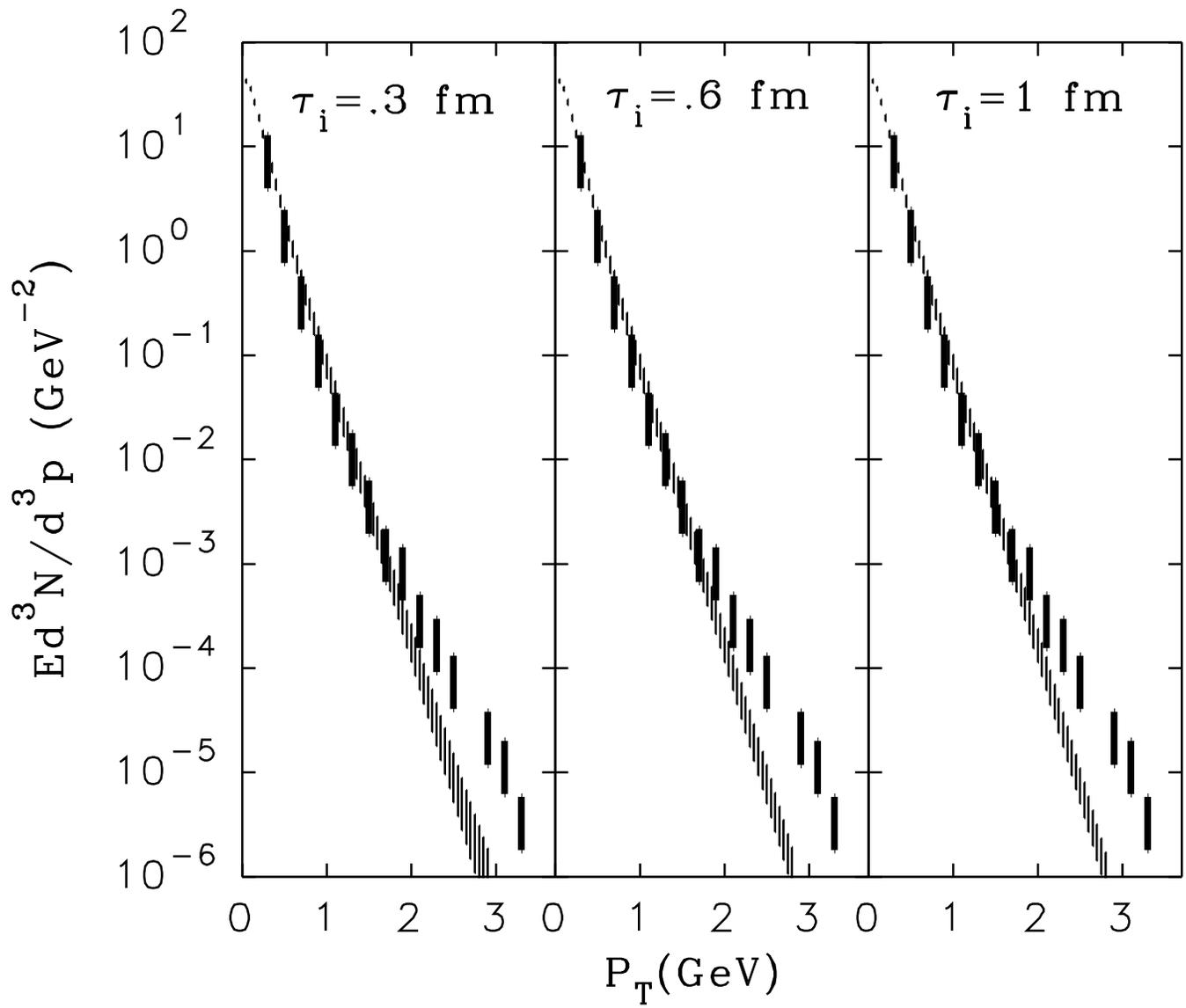
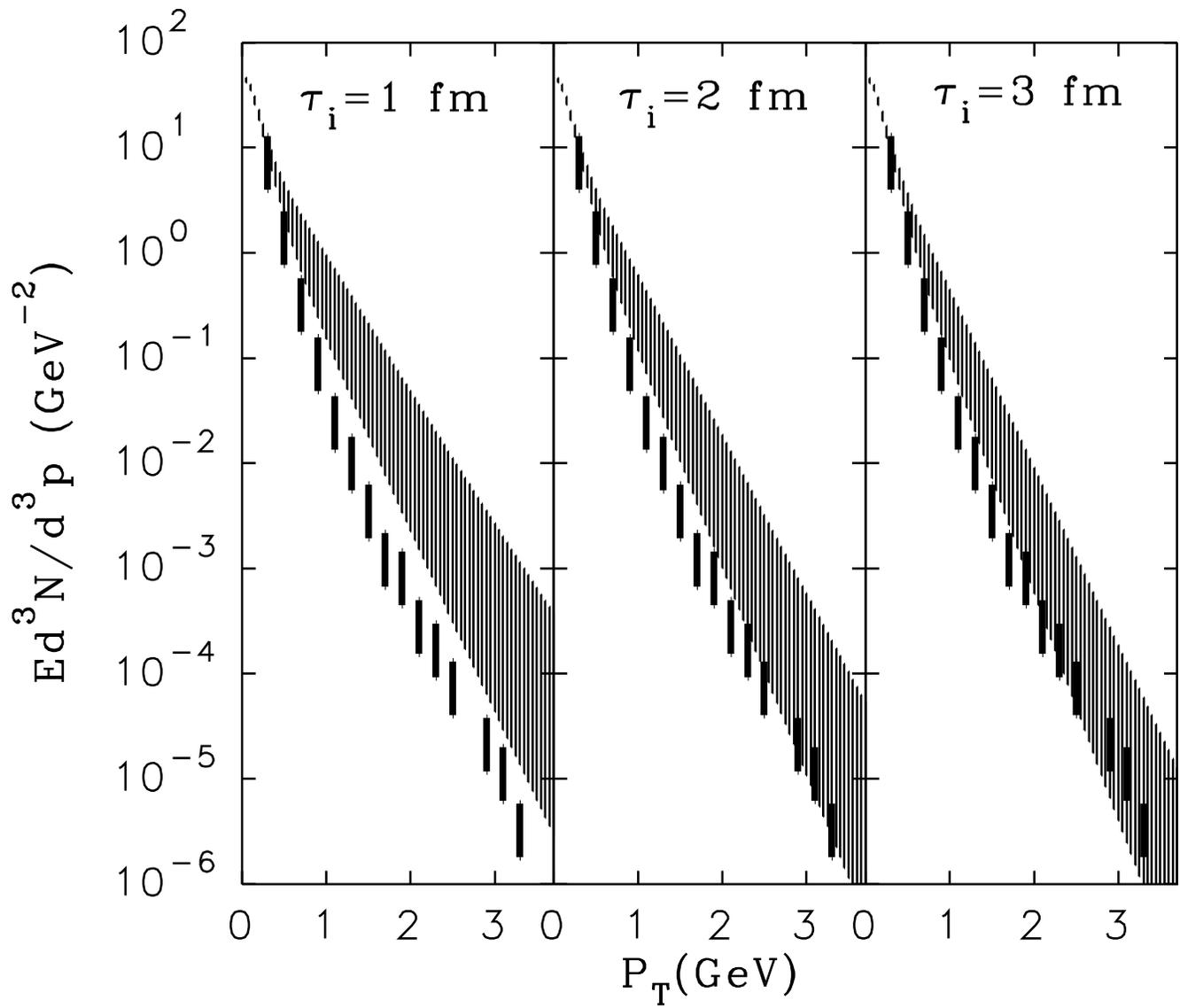
\end{document}